\documentclass[reprint,amsmath,amssymb,aip,?rl]{revtex4-1}
\usepackage[english] {babel}
\usepackage{hyperref}
\usepackage{amsfonts,graphicx}
\usepackage{hyperref}
\usepackage{epstopdf}

\usepackage{dcolumn}
\usepackage{bm}

\usepackage{siunitx}
\usepackage[caption=false]{subfig}

\begin{document}


\title{Mean Transverse Energy of Ultrananocrystalline Diamond Photocathode}

\author{Gongxiaohui Chen}
 \email{gchen26@hawk.iit.edu}
 \affiliation{Department of Physics, Illinois Institute of Technology, Chicago, IL, 60616}
 
\author{Gowri Adhikari}
 \affiliation{Department of Physics, University of Illinois at Chicago, Chicago, IL, 60607}

\author{Linda Spentzious}
 \affiliation{Department of Physics, Illinois Institute of Technology, Chicago, IL, 60616}
 
\author{Kiran Kumar Kovi}
 \affiliation{Euclid TechLabs LLC, Bolingbrook, IL, 60440}
\author{Sergey Antipov}
 \affiliation{Euclid TechLabs LLC, Bolingbrook, IL, 60440}
\author{Chunguang Jing}
 \affiliation{Euclid TechLabs LLC, Bolingbrook, IL, 60440}
 
 \author{W. Andreas Schroeder}
 \email{andreas@uic.edu}
 \affiliation{Department of Physics, University of Illinois at Chicago, Chicago, IL, 60607}
 
\author{Sergey V. Baryshev}
\email{serbar@msu.edu}
 \affiliation{Department of Electrical and Computer Engineering, Michigan State University, East Lansing, MI, 48824}


\begin{abstract}
Nitrogen incorporated ultrananocrystalline diamond ((N)UNCD) could be an enabling material platform for photocathode applications due to its high emissivity. While the quantum efficiency (QE) of UNCD was reported by many groups, no experimental measurements of the intrinsic emittance/mean transverse energy (MTE) have been reported. Here, MTE measurement results for an (N)UNCD photocathode in the photon energy range of 4.41 to 5.26 eV are described. The MTE demonstrates no noticeable dependence on the photon energy, with an average value of 266 meV. This spectral behavior is shown to not to be dependent upon physical or chemical surface roughness and inconsistent with low electron effective mass emission from graphitic grain boundaries, but may be associated with emission from spatially-confined states in the graphite regions between the diamond grains. The combined effect of fast-growing QE and constant MTE with respect to the excess laser energy may pave the way to bright UNCD photocathodes.
\end{abstract}

\maketitle

Photocathode-based RF and pulsed DC guns are bright electron injectors for free electron lasers and advanced time resolved microscopes \cite{dowell_cathode_2010}. Further progress of electron laser and microscopy facilities (improved sensitivity, spatiotemporal resolution, high throughput) largely depends on development and understanding of materials with the potential to be utilized as photocathodes. Photocathode development challenges include achieving simultaneously ($i$) high QE, ($ii$) high transverse coherence (meaning low intrinsic emittance/low MTE), ($iii$) rapid response time.

The ratio of the charge to the MTE determines the photocathode brightness, which in many applications is the most critical figure of merit. For a classical metal photocathode such as copper, the Fowler-Dubridge law \cite{dubridge_theory_1933} predicts that the emitted charge is a fast-growing function of excess energy (a power law), where excess energy  $\Delta E$ is the difference between the laser primary incident photon energy $\hbar \omega$ and the work function $\phi$ defined as $\Delta E=\hbar \omega - \phi$. Dowell and Schmerge \cite{dowell_quantum_2009} have found that the transverse momentum for metals also grows with excess energy as $\sim \sqrt{\hbar \omega - \phi}$. For the latter reason, to attain the highest quality (low divergence) electron beam metal photocathodes are often operated in the near threshold region (having the smallest $\Delta E$, with the primary photon energy nearly matching the work function), although brightness increases with excess energy.

A great number of metal and thin film alkali antimonide photocathodes obey the Dowell-Schmerge (DS) model \cite{dowell_quantum_2009, bazarov_thermal_2011, feng_novel_2015}. However, some semiconductor photocathodes, e.g. GaAs and PbTe, show various MTE versus excess energy trends that are different from those specific to metals. Negative electron affinity (NEA) GaAs photocathodes \cite{karkare_ultrabright_2014}, for instance, demonstrate $\sim$1,000-fold QE increase as the excess energy increases from 0 to about 1 eV while the MTE remains low and nearly constant with the same $\Delta E$ range (within measurement precision).

(N)UNCD is another example of a NEA photocathode that has high electron conductivity through the bulk of a semi-metallic nature. The NEA is induced via surface C-H dipole formation when UNCD is processed in a hydrogen plasma (UNCD:H) \cite{cui_electron_1998-1, sun_combined_2011-1}. Compared to cesiated GaAs, the C- H dipole is stable in air. The resulting QE is high ($\sim$$10^{-3}$) \cite{perez_quintero_high_2014, ohmagari_carrier_2014, mazellier_photoemission_2014} and could potentially be further increased by choosing a different $n$-type dopant such as phosphorous \cite{koeck_thermionic_2009}. The intrinsic as-grown surface roughness is low, less than 10 nm. The low physical roughness suggests that the beam emittance can be low. In order to elucidate the ultimate performance of $n$-type UNCD, the transverse electron momentum $\Delta  p_T$ is a fundamental parameter to determine (or equivalently the MTE, since MTE=$\Delta p_T^2/2m$, with $m$ being the electron mass).

In this paper, the MTE of a (N)UNCD photocathode was experimentally measured over an excess energy range of 1 eV. It was found that the MTE does not noticeably depend on the excess energy with an average value of 266 meV. It is proposed that this spectral behavior is due to emission from spatially-confined states in the graphite regions (i.e. grain boundaries) between the diamond grains.

\begin{figure}
	\centering
	\subfloat{\label{subfig:Raman_UIC}\includegraphics[width=0.4\textwidth]{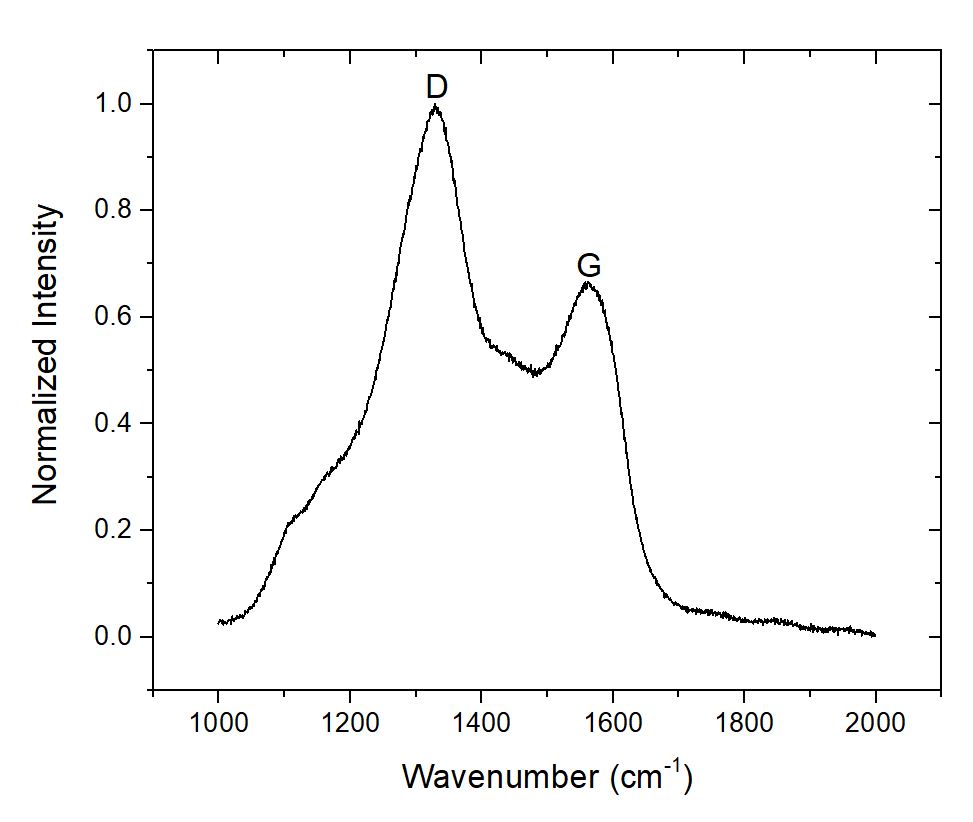}}
	
	\subfloat{\label{subfig:emittance_test_stand}\includegraphics[width=0.45\textwidth]{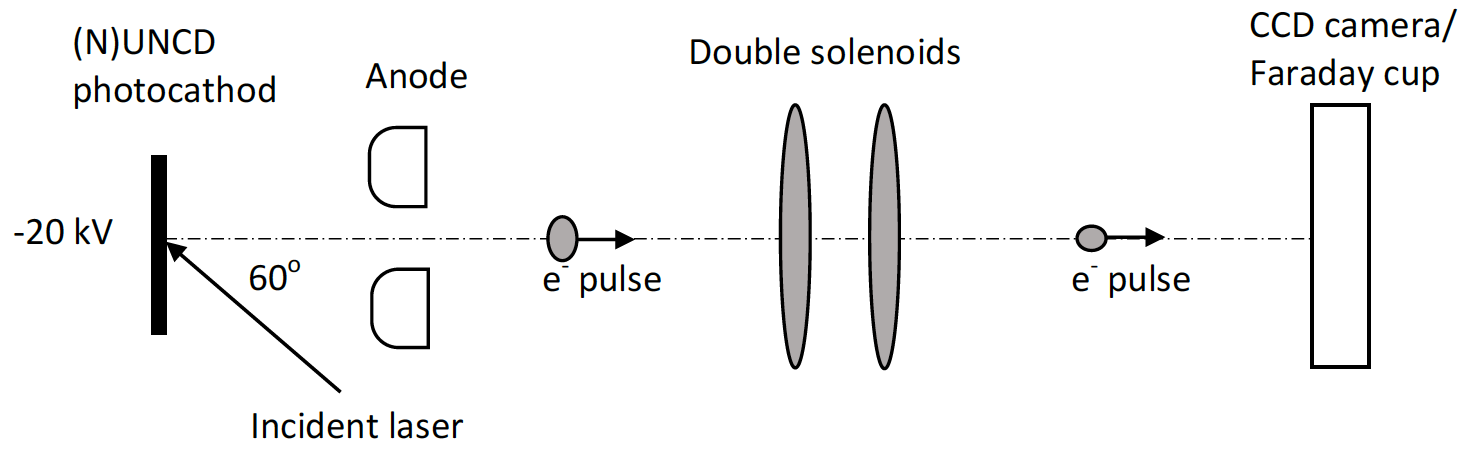}}
\caption{(a) Raman spectrum of the (N)UNCD sample showing the characteristic diamond (D) and graphite (G) peaks; (b) The solenoid scan system  for MTE measurement. Electrons accelerated in the DC gun travel through the double-solenoid lens and imping on the scintillator screen imaged by a CCD camera.}
\end{figure}

The (N)UNCD film, 160 nm thick, was deposited on highly doped $n$-Si substrate by a microwave plasma chemical vapor deposition (MPCVD) method. Deposition parameters were identical to those reported in the paper of Pérez Quintero \textit{et al.} \cite{perez_quintero_high_2014}. A Raman spectrum confirming the (N)UNCD chemical bonding structure is presented in Fig. \ref{subfig:Raman_UIC}. The work function of (N)UNCD film was found to be 4.2 eV using a Kelvin probe.

\begin{figure}
\includegraphics[width=0.35\textwidth]{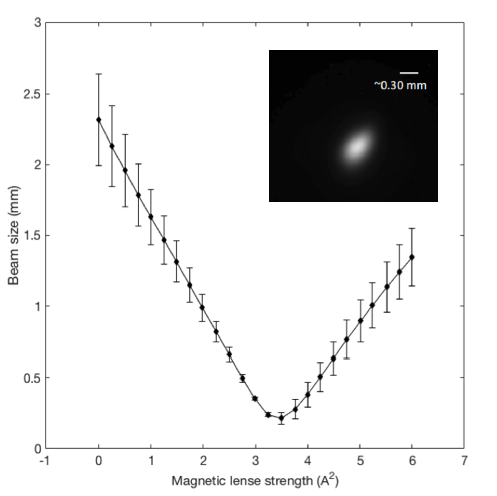}
\caption{Solenoid scan data for an incident 4.75 eV photon energy. The top and bottom of the ‘error bars’ represent the measured horizontal and vertical $HWe^{-1}M$ beam sizes, respectively, and the connected dots the average beam size. The inset shows the actual beam image at the focal point for the solenoid scan, i.e. at a solenoid current of 1.83 A.}
\label{fig:beam_size_beam_img}
\end{figure}

The experimental setup for measuring the MTE of the emitted electrons using the solenoid scan technique \cite{graves_measurement_2001} is shown in Fig \ref{subfig:emittance_test_stand}. The electron pulses are generated in a 20 kV DC gun using tunable ultraviolet (UV) radiation from a 30 MHz repetition rate sub-picosecond laser system driven by a diode pumped and mode-locked Yb:KGW oscillator \cite{berger_high-power_2008}. Briefly, the 1046 nm, 0.25 ps pulse duration output from the 2 W Yb:KGW oscillator is used to generate a continuum in a photonic crystal fiber which is then selectively amplified by optical parametric amplification (OPA) in lithium triborate (LBO) nonlinear crystals and the resulting signal and idler pulses are subsequently sum frequency mixed with the second and third harmonics of the Yb:KGW laser to generate tunable UV radiation. Together with the 3.56 eV (349 nm) third and 4.75 eV (262 nm) fourth harmonics of the Yb:KGW laser, this provides a UV radiation source with almost continuous 3.0-5.3 eV (235-410 nm) tunability. The $p$-polarized UV laser beam is circular for the third and fourth harmonics and elliptical (aspect ratio $\sim$1:1.4) from the sum frequency generation. At the employed 60$\si{\degree}$ incidence angle, the measured half-width 1/e maximum, dubbed as $HWe^{-1}M,$ ($x, y$) irradiance spot sizes on the photocathode surface are (220, 80)$\pm$5 $\mu$m for the sum frequency generated UV radiation and (240, 120)$\pm$5 $\mu$m for the fourth harmonic at 4.75 eV. The known electron source size then provides a required input parameter for the extended analytical Gaussian (AG) simulation \cite{michalik_analytic_2006, berger_semianalytic_2010} of the electron beam propagation from emission to detection, with a Ce:GAGG scintillation crystal and a CCD camera, through the solenoid scan MTE measurement system. Fig. \ref{fig:beam_size_beam_img} displays the measured $HWe^{-1}M$ electron beam sizes at the scintillator as a function of the square of the current (i.e., focal strength) passing through the two solenoid lenses (counter wound to avoid image rotation effects) for the 4.75 eV incident UV photon energy. In this case, the AG model simulation fit using either the horizontal ($x$) or vertical ($y$) spot sizes (range indicated by the ‘error bar span’ in Fig. \ref{fig:beam_size_beam_img}), or the average beam size (connected dots in Fig. \ref{fig:beam_size_beam_img}), generates an MTE value for the emitted electrons of 290($\pm$40) meV. We note that the measurements were made in the low charge regime so that the beam imaging and MTE calculations were not affected by space charge. Additionally, the low 0.45 MV/m acceleration field gradient on the (N)UNCD photocathode surface in the DC gun ensured that dark current effects were negligible and produced a Schottky effect of only $\sim$30 meV that is comparable to the thermal energy at room temperature.

\begin{figure}
\includegraphics[width=0.4\textwidth]{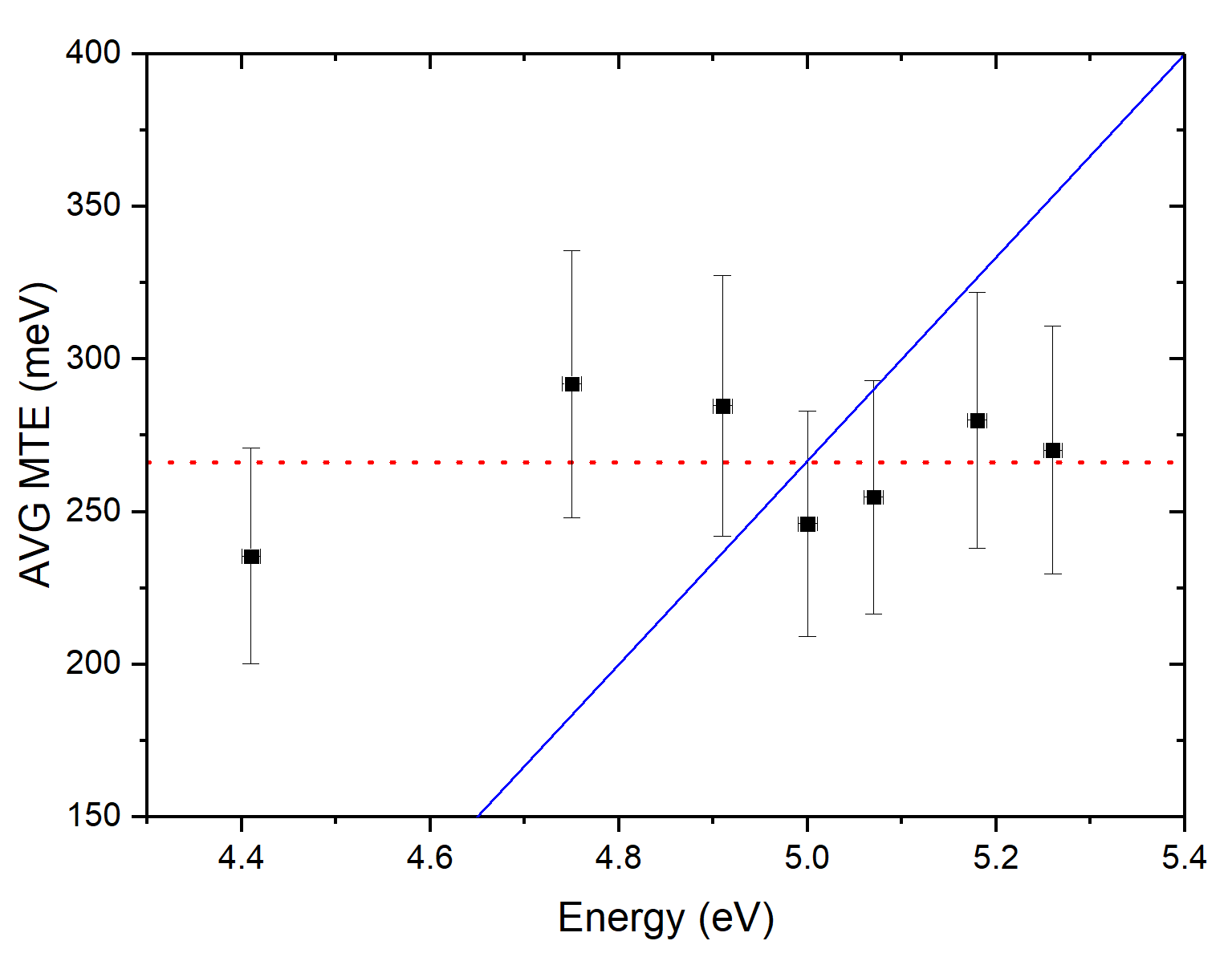}
\caption{MTE values extracted from the solenoid scan data plotted versus the incident photon energy. The red dotted line depicts the mean MTE equal to 266 meV. The blue line is the DS model for which $(\hbar \omega - \phi)/3$.}
\label{fig:MTE_simple_model}
\end{figure}

Fig. \ref{fig:MTE_simple_model} represents the full summary of MTE values obtained through multiple solenoid scans performed at multiple primary laser photon energies. The measured MTE values display a near flat trend, remaining independent of the UV photon energy in the range of 4.41 to 5.26 eV. The red dashed line in Fig. \ref{fig:MTE_simple_model} represents the mean MTE value of 266 meV. Such a flat response is not common for most photocathode systems. To compare the deduced MTE($\hbar \omega$) dependence of (N)UNCD against the DS model, $(\hbar \omega - \phi)/3$ (blue solid line) is plotted using $\phi$=4.2 eV. The drastic deviation from the metallic photoemission model calls for an alternative scenario.

(N)UNCD is a graphitic rich two-phase material that consists of $sp^3$ diamond grains and $sp^2$ graphitic grain boundaries. It has been reported that electron emission (photo- or field- emission) in this and related materials preferentially originates from the grain boundaries \cite{harniman_direct_2015, kurian_role_2014}.  Accordingly, a possible explanation for the observed spectral dependence of the MTE is physical and chemical roughness associated with the nano-granularity of the photocathode material. However, the analysis presented in the paper of Karkare \textit{et al.} \cite{karkare_effects_2015} suggests that neither play a role in our case. The low $\sim$0.5 MV/m surface acceleration field in the 20 kV DC gun excludes any significant increase in MTE due to the 10 nm surface roughness due to the UNCD grain size. Similarly, effects on the MTE due to chemical roughness are negligible despite the difference in diamond $sp^3$ and graphitic $sp^2$ grain boundary work functions, which can be as large as 1.5 eV. The latter is due to the rapid decay from the photocathode surface of transverse field modulations due to work function variations for the small $\sim$10nm-scale (grain size) periodicity.  

\begin{figure}
        \subfloat{\label{subfig:e_like_heavy_fff}\includegraphics[height=1.2in]{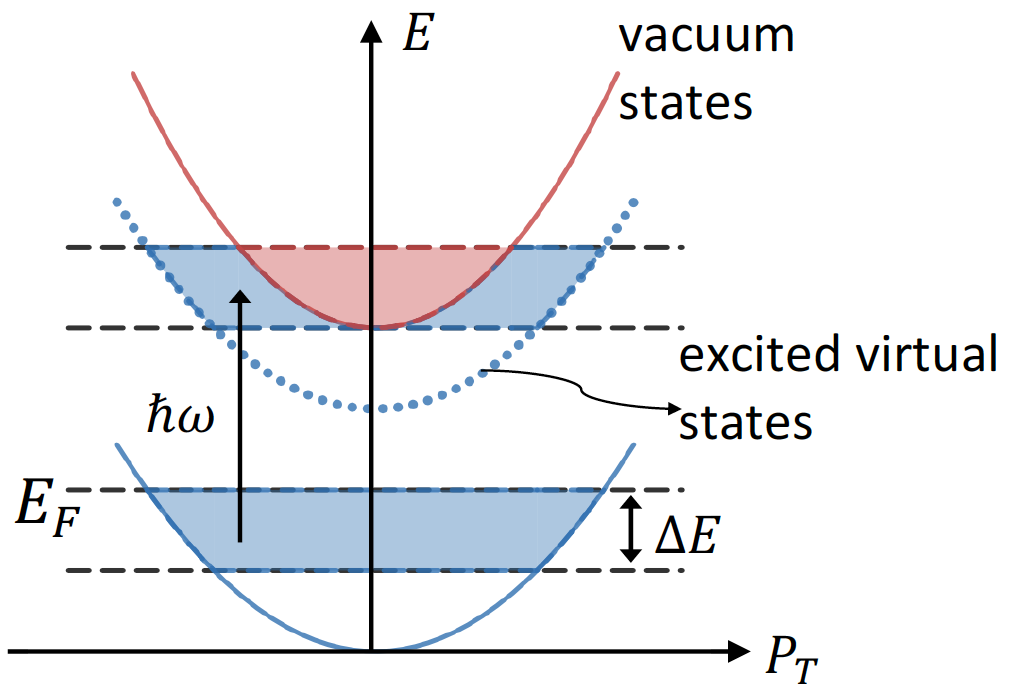}}
        \subfloat{\label{subfig:e_like_light_fff}\includegraphics[height=1.2in]{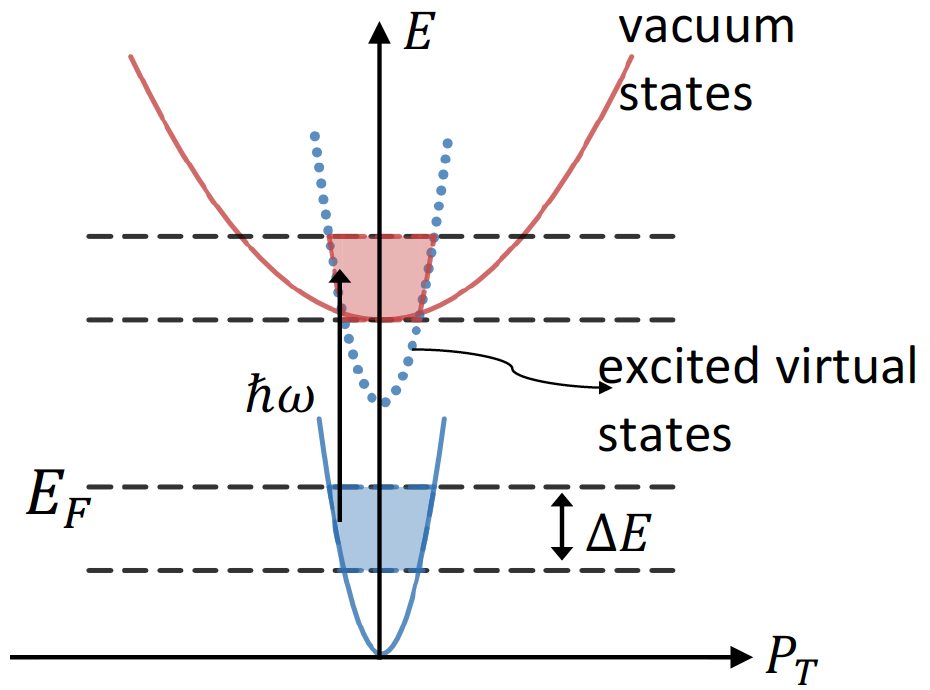}}
    \caption{Simplified one-step photoemission (energy versus transverse momentum) diagrams that capture the effective mass effect on MTE: (a) photoemission from states with dispersion corresponding to free electron mass $m_0$ resulting in DS like behavior; (b) photoemission from states with a small effective electron mass $m^*$ resulting in MTE behavior insensitive to the excess energy. Solid blue lines represent the actual metallic-like parabolic $\varepsilon$$(k)$ dispersion relations, where the blue highlighted area indicates states with sufficient excess energy for photoemission. The dotted blue line represents the virtual excited states, and the red highlighted area depicts the final allowed photo-emitting electron states.}
\label{fig:e_like}
\end{figure}

Studies of the bulk electron transport in (N)UNCD have confirmed that electrons percolate through grain boundary networks with an effective mass equivalent to that in graphite \cite{chen_using_2012}, i.e. an electron effective mass 1/18 of the electron rest mass $m_0$. Such a small electron mass in turn suggests narrow electronic energy bands, which could affect the MTE of electron emission. Consider a one-step quantum mechanical photoemission mechanism where the incident photon momentum is considered negligible, so that the excited virtual states (intermediate states between the electron states inside the material and the free space electron states) have the same dispersion relation $\varepsilon$$(k)$ as in the initial state, i.e. as inside the material before photon absorption. Fig. \ref{fig:e_like} compares such a one-step photoemission process from states with a dispersion determined by the rest electron mass $m_0$ (a) to one from states with a small effective mass $m^*$ (b). With transverse momentum conservation in photoemission, it is clear that a small effective mass for the emitting states can serve to restrict the transverse momentum, and hence the MTE, of the photo-emitted electrons (Fig. \ref{subfig:e_like_light_fff}). In contrast, the transverse momentum for an electron emitted from a ‘perfect’ metal photocathode with a free electron mass dispersion is only restricted by the vacuum state dispersion (Fig. \ref{subfig:e_like_heavy_fff}), resulting in a MTE($\hbar \omega$) variation in accordance with the DS model. A rule-of-thumb proposed in the paper of Rickman \textit{et al.} \cite{rickman_intrinsic_2013} is that transverse electron momentum $p_T$ in the photoemission process that is actually realized in experiment is restricted by either $\sqrt{2m_0(\hbar \omega - \phi)}$ or $\sqrt{2m^*E_F}$ (where $E_F$ is the Fermi energy), whichever is smaller. Since for graphite $m^*=0.045m_0$ \cite{tatar_electronic_1982} and $E_F$ may be as small as 30 meV \cite{schneider_electronic_2010}, the product $m^* E_F$ is always less than the product $m_0 \Delta E$. From this simplified consideration, the MTE should be dependent upon the Fermi energy, but the measured quantity is almost an order of magnitude larger. 

An alternative explanation for the measured relatively invariant MTE of above threshold electron emission (Fig. \ref{fig:MTE_simple_model}), and consistent with the above, is that the emitted electrons originate from spatially-confined states in the graphite regions between the diamond grains.  If so, transverse momentum conservation in photoemission implies that the MTE should reflect the rms momentum (i.e., size in momentum space) of these states, assuming surface effects can be neglected.  To obtain the observed MTE of $\sim$270 meV, the emitting states would then need to be confined to a spatial region of $\sim$1 nm – this is obtained through the Heisenberg's uncertainty principle. The emitter size of $\sim$1 nm is well comparable with the grain boundary size in (N)UNCD. Further experimental and theoretical investigations will be needed to shed light on the exact photoemission mechanism and hence the realized MTE values: For example, atomic resolution EELS electron microscopy of the graphite grain boundaries and detailed band structure studies of (N)UNCD.

In summary, we reported measurement results of MTE for a (N)UNCD photocathode. No noticeable dependence of the MTE on the excess energy over a range of 1 eV was measured, which is non-conventional behavior that has been observed so far in only a few photocathode systems. This spectral dependence is shown to not to be dependent upon surface roughness (physical or chemical) and inconsistent with low electron effective mass emission from graphitic grain boundaries, but is likely to be associated with emission from spatially-confined states in the $\sim$1 nm graphite regions between the diamond grains. Given this promising intrinsic emission property of (N)UNCD, the next step would be to measure the MTE and QE of a hydrogen surface terminated (N)UNCD sample. If the MTE remains nearly constant, while the QE increases as expected due to the NEA produced after the hydrogen termination, then the (N)UNCD material has great potential as a next generation photocathode, given its additional ability to withstand poor vacuum without significant degradation.

\begin{acknowledgments}
This project is supported by NSF grant No. NSF-1739150, DOE SBIR program grant No. DE-SC0013145 and NSF grant No. PHYS-1535279. Use of the Center for Nanoscale Materials, an Office of Science user facility, was supported by the U.S. Department of Energy, Office of Science, Office of Basic Energy Sciences, under Contract No. DE-AC02-06CH11357. S.V.B. was supported by funding from the College of Engineering, Michigan State University, under Global Impact Initiative.
\end{acknowledgments}

\bibliography{refff}

\end{document}